\documentclass[aos]{imsart}
\setattribute{journal}{name}{}
\usepackage{amsmath}
\usepackage{amsbsy}
\usepackage{graphicx,amsmath,amssymb,fullpage,amsfonts,bbm,bbold}
\usepackage{amsthm}
\usepackage{epsfig}
\usepackage{color}

\usepackage{booktabs}
\usepackage{enumerate}

\DeclareMathOperator{\iid}{\stackrel{\mbox{\tiny iid} }{\sim}}

\newcommand{\X}{\mathbf{X}}
\newcommand{\Y}{\tilde{Y}}

\newcommand{\N}{\mbox{{\small\textsc{N}}}}

\newcommand{\C}{\mbox{{\small\textsc{C}}}}

\newcommand{\E}{\mbox{E}}

\newcommand{\I}{\mathbf{I}}

\usepackage{bm}
\usepackage{natbib}
\begin{document}
\begin{frontmatter}
\title{Decoupling shrinkage and selection in Bayesian \\linear models: a posterior summary perspective}
\author{P. Richard Hahn} \and \author{Carlos M. Carvalho}
\affiliation{Booth School of Business and McCombs School of Business}
\date{}

\begin{abstract}
Selecting a subset of variables for linear models remains an active area of research.  This paper reviews many of the recent contributions to the Bayesian model selection and shrinkage prior literature.   A posterior variable selection summary is proposed,  which distills a full posterior distribution over regression coefficients into a sequence of sparse linear predictors.  
\end{abstract}


\end{frontmatter}



\section{Introduction}
This paper revisits the venerable problem of variable selection in linear models. The vantage point throughout is Bayesian:  a  normal likelihood is assumed and inferences are based on the posterior distribution, which is arrived at by conditioning on observed data. 


In applied regression analysis, a ``high-dimensional" linear model can be one which involves tens or hundreds of variables, especially when seeking to compute a full Bayesian posterior distribution.  Our review will be from the perspective of a data analyst facing a problem in this ``moderate" regime.  Likewise, we focus on the situation where the number of predictor variables, $p$, is fixed.
 
 In contrast to other recent papers surveying the large body of literature on Bayesian variable selection \citep{Liangetal08, Berger12} and shrinkage priors \citep{o2009review, Polson2012local}, our review focuses specifically on the  relationship between variable selection priors and shrinkage priors.  Selection priors and shrinkage priors are related both by the statistical ends they attempt to serve (e.g., strong regularization and efficient estimation) and also in the technical means they use to achieve these goals (hierarchical priors with local scale parameters).  We also compare these approaches on computational considerations.  
 
 Finally, we turn to variable selection as a problem of posterior summarization.  We argue that if variable selection is desired primarily for parsimonious communication of linear trends in the data, that this can be accomplished as a post-inference operation irrespective of the choice of prior distribution.  To this end, we introduce a posterior variable selection summary, which distills a full posterior distribution over regression coefficients into a sequence of sparse linear predictors.  In this sense ``shrinkage" is decoupled from ``selection".   

We begin by describing the two most common approaches to this scenario and show how the two approaches can be seen as special cases of an encompassing formalism.

\subsection{Bayesian model selection formalism}
A now-canonical way to formalize variable selection in Bayesian linear models is as follows.  Let $\mathcal{M}_{\phi}$ denote a normal linear regression model indexed by a vector of binary indicators $\phi = (\phi_1,\dots,\phi_p) \in \{0,1 \}^p$ signifying which predictors are included in the regression. Model $\mathcal{M}_{\phi}$ defines the data distribution as
\begin{equation}
(Y_i | \mathcal{M}_{\phi},\beta_{\phi}, \sigma^2 ) \sim \N(X^{\phi}_i \beta_{\phi},\sigma^2)
\end{equation}  
where $X^{\phi}_i$ represents the $p_{\phi}$-vector of predictors in model $\mathcal{M}_{\phi}$. Given a sample $\mathbf{Y}=(Y_1,\dots,Y_n)$ and prior $\pi(\beta_{\phi},\sigma^2)$, the inferential target is the set of posterior model probabilities defined by
\begin{equation}
p(\mathcal{M}_{\phi} \mid \mathbf{Y}) = \frac{p(\mathbf{Y} \mid \mathcal{M}_{\phi}) p(\mathcal{M}_{\phi})}{\sum_{\phi} p(\mathbf{Y} \mid \mathcal{M}_{\phi})p(\mathcal{M}_{\phi})},
\end{equation}
where $p(\mathbf{Y} \mid \mathcal{M}_{\phi}) = \int p(\mathbf{Y} \mid \mathcal{M}_{\phi},\beta_{\phi},\sigma^2)\pi(\beta_{\phi}, \sigma^2) d \beta_{\phi} d\sigma^2$ is the marginal likelihood of model $\mathcal{M}_\phi$ and $p(\mathcal{M}_{\phi})$ is the prior over models.   

Posterior inferences concerning a quantity of interest $\Delta$ are obtained via Bayesian model averaging (or BMA), which entails integrating over the model space 
\begin{equation}\label{BMA}
p(\Delta \mid \mathbf{Y}) = \sum_{\phi} p(\Delta \mid \mathcal{M}_{\phi}, \mathbf{Y}) p(\mathcal{M}_{\phi} \mid \mathbf{Y}).
\end{equation}
As an example, optimal predictions of future values of $\Y$ under squared-error loss are defined through
\begin{equation} \label{MMmean}
\E(\Y \mid \mathbf{Y}) \equiv \sum_{\phi} \E(\Y \mid \mathcal{M}_{\phi}, \mathbf{Y}) p(\mathcal{M}_{\phi}\mid \mathbf{Y}).
\end{equation}
An early reference adopting this formulation is \cite{Raftery97}; see also \cite{Clyde04}. 

Despite its straightforwardness, carrying out variable selection in this framework demands attention to detail: priors over model-specific parameters must be specified, priors over models must be chosen, marginal likelihood calculations must be performed and a $2^p$-dimensional discrete space must be explored. These concerns have animated Bayesian research in linear model variable selection for the past two decades.  

Regarding model parameters, the consensus default prior for  model parameters is $\pi(\beta_{\phi}, \sigma^2) = \pi(\beta \mid \sigma^2) \pi(\sigma^2) = \N(0, g\Omega) \times \sigma^{-1}$.  The most widely-studied choice of prior covariance is  $\Omega = \sigma^2(\X_{\phi}^t\X_{\phi})^{-1}$, referred to as ``Zellner's $g$-prior" \citep{Zellner}, a ``$g$-type" prior or simply $g$-prior. Notice that this choice of $\Omega$ dictates that the prior and likelihood are conjugate normal-inverse-gamma pairs (for a fixed value of $g$). 

For reasons detailed in \cite{Liangetal08}, it is advised to place a prior on $g$ rather than use a fixed value.  Several recent papers describe  priors $p(g)$ that still lead to efficient computations of marginal likelihoods; see \cite{Liangetal08}, \cite{maruyamaGeorge2011}, and \cite{Berger12}.  Each of these papers (as well as the earlier literature cited therein) study priors of the form
 \begin{equation}
 p(g) \propto g^d (g+ b)^{-(a+c+d +1)}
 \end{equation}  
 with $a > 0$, $b > 0$, $c>-1$, and $d > -1$.  (The support of $g$ will be lower bounded by a function of the hyper parameter $b$.)  Specific configurations of these hyper parameters recommended in the literature include: $\lbrace a=1, b=1, d=0 \rbrace$ \citep{cui2008empirical}, $\lbrace a=1/2, b=1\;\; (b=n), c=0, d=0 \rbrace$ \citep{Liangetal08},  and  $\lbrace c = -3/4, d=(n-5)/2 - p_{\phi}/2 + 3/4 \rbrace$ \citep{maruyamaGeorge2011}.
 
\cite{Berger12} motivates the use of such priors from a formal testing perspective, using a variety of intuitive desiderata.  Regarding prior model probabilities see \cite{ScottBerger2010}, who recommend a hierarchical prior of the form $\phi_j \iid \mbox{Ber}(q)$,  $q \sim \mbox{Unif}(0,1)$.


\subsection{Shrinkage regularization priors}
Although the formulation above provides a valuable theoretical framework, it does not necessarily represent an applied statistician's first choice.  To assess which variables contribute dominantly to trends in the data, the goal may be simply to mitigate---rather than categorize---spurious correlations.  Thus, faced with many potentially irrelevant predictor variables, a common first choice would be a powerful {\em regularization prior}. 

Regularization --- understood here as the intentional biasing of an estimate to stabilize posterior inference --- is inherent to most Bayesian estimators via the use of proper prior distributions and is one of the often-cited advantages of the Bayesian approach.  More specifically, regularization priors refer to priors explicitly designed with a strong bias for the purpose of separating reliable from spurious patterns in the data.  In linear models, this strategy takes the form of zero-centered priors with sharp modes and simultaneously fat tails.

A well-studied class of priors fitting this description will serve to connect continuous priors to the model selection priors described above.  {\em Local scale mixture} of normal distributions are of the form
\begin{equation}\label{local}
\pi(\beta_j \mid \lambda) = \int{\mbox{N}(\beta_j \mid 0,\lambda^2 \lambda_j^2)\pi(\lambda_j^2)} d \lambda_j,
\end{equation}
where different priors are derived from different choices for $\pi(\lambda_j^2)$.  

The last several years have seen tremendous interest in this area, motivated by an analogy with penalized-likelihood methods  \citep{lasso96}.  Penalized likelihood methods with an additive penalty term lead to estimating equations of the form
\begin{equation}\label{ple}
\sum_i h(Y_i, \X_i, \beta) + \alpha Q(\beta)
\end{equation}
where $h$ and $Q$ are positive functions and their sum is to be minimized; $\alpha$ is a scalar tuning variable dictating the strength of the penalty.  Typically, $h$ is interpreted as a negative log-likelihood, given data $\mathbf{Y}$, and $Q$ is a penalty term introduced to stabilize maximum likelihood estimation.  A common choice is  $Q(\beta) = ||\beta||_1$, which yields sparse optimal solutions $\beta^*$ and admits fast computation \citep{lasso96}; this choice underpins the {\em  lasso} estimator, a mnemonic for ``least absolute shrinkage and selection operator".

\cite{ParkCasella} and \cite{HansLasso09} ``Bayesified" these expressions by interpreting $Q(\beta)$ as the negative log prior density and developing algorithms for sampling from the resulting Bayesian posterior, building upon work of earlier Bayesian authors \citep{spiegelhalter1977test,west1987scale,pericchi1991robust,pericchi1992exact}.  Specifically, an exponential prior $\pi(\lambda_j^2) = \mbox{Exp}(\alpha^2)$ leads to independent Laplace (double-exponential) priors on the $\beta_j$, mirroring expression (\ref{ple}).

This approach has two implications unique to the Bayesian paradigm.  First, it presented an opportunity to treat the global scale parameter $\lambda$ (equivalently the regularization penalty parameter $\alpha$) as a hyper parameter to be estimated.  Averaging over $\lambda$ in the Bayesian paradigm has been empirically observed to give better prediction performance than cross-validated selection of $\alpha$ (e.g., \cite{HansLasso09}).  Second, a Bayesian approach necessitates forming point estimators from posterior distributions; typically the posterior mean is adopted on the basis that it minimizes mean squared prediction error.  Note that posterior mean regression coefficient vectors from these models are non-sparse with probability one.  Ironically,  the two main appeals of the penalized likelihood methods---efficient computation and sparse solution vectors $\beta^*$---were lost in the migration to a Bayesian approach.

Nonetheless, wide interest in ``Bayesian lasso" models paved the way for more general local shrinkage regularization priors of the form (\ref{local}).  In particular,  \cite{Horseshoe10} develops a  prior over location parameters that attempts to shrink irrelevant signals strongly toward zero while avoiding excessive shrinkage of relevant signals.  To contextualize this aim, recall that solutions to $\ell_1$ penalized likelihood problems are often interpreted as (convex) approximations to more challenging formulations based on $\ell_0$ penalties.  As such, it was observed that the global $\ell_1$ penalty ``overshrinks" what ought to be large magnitude coefficients.  The \cite{Horseshoe10} prior may be written as
\begin{equation}
\begin{split}
\pi(\beta_j \mid \lambda)& = \N(0,\lambda^2 \lambda_j^2),\\
 \lambda_j &\iid \C^+(0,1).
 \end{split}
\end{equation}
with $ \lambda \sim \C^+(0,1)$ or   $ \lambda \sim \C^+(0,\sigma^2)$.  The choice of half-Cauchy arises from the insight that for scalar observations $y_j \sim \N(\theta_j, 1)$ and prior $\theta_j \sim \N(0, \lambda_j^2)$, the posterior mean of $\theta_j$ may be expressed:
\begin{equation}
\E(\theta_j\mid y_j) = \{1-\E(\kappa_j \mid y_j)\} y_j,
\end{equation}
where $\kappa_j = 1/(1+\lambda_j^2)$. The authors observe that U-shaped Beta(1/2,1/2) distributions (like a horseshoe) on $\kappa_j$ imply a prior over $\theta_j$ with high mass around the origin but with polynomial tails.  That is, the ``horseshoe" prior encodes the assumption that some coefficients will be very large and many others will be very nearly zero.  This U-shaped prior on $\kappa_j$ implies the half-cauchy prior density $\pi(\lambda_j)$.  The implied prior on $\beta$ has Cauchy-like tails and a pole at the origin which entails  more aggressive shrinkage than a Laplace prior.

Other choices  of $\pi(\lambda_j)$ lead to different ``shrinkage profiles" on the ``$\kappa$ scale".   \cite{Polson2012local} provides an excellent taxonomy of the various priors over $\beta$ that can be obtained as scale-mixtures of normals.  The horseshoe and similar priors (e.g., \cite{Griffin12}) have proven empirically to be fine default choices for regression coefficients:  they lack hyper parameters, forcefully separate strong from weak predictors, and exhibit robust predictive performance.  




\subsection{Model selection priors as shrinkage priors}
It is possible to express model selection priors as shrinkage priors.  To motivate this re-framing, observe that the posterior mean regression coefficient vector is not well-defined in the model selection framework. Using the model-averaging notion, the posterior average $\beta$ may be be {\em defined} as:
\begin{equation}\label{modelav}
\E(\beta \mid \mathbf{Y}) \equiv \sum_{\phi} \E(\beta \mid \mathcal{M}_{\phi}, \mathbf{Y}) p(\mathcal{M}_{\phi} \mid \mathbf{Y}),
\end{equation}
 where $\E(\beta_j \mid \mathcal{M}_{\phi}, \mathbf{Y}) \equiv 0$ whenever $\phi_j = 0$. Without this definition, the posterior expectation of $\beta_j$ is undefined in models where the $j$th predictor does not appear.  More specifically, as the likelihood is constant in variable $j$ in such models, the posterior remains whatever the prior was chosen to be. 
 
 To fully resolve this indeterminacy, it is common to set $\beta_j$ identically equal to zero in models where the $j$th predictor does not appear, consistent with  the interpretation that  $\beta_j \equiv \partial \E(Y)/\partial X_j$.  A hierarchical prior reflecting this choice may be expressed
 \begin{equation}\label{eq:spikeslab}
 \pi(\beta \mid \sigma^2, \phi) = \N(0, g \Lambda \Omega \Lambda^t)
 \end{equation}
 where $\Lambda \equiv \mbox{diag}((\lambda_1, \lambda_2, \dots, \lambda_p))$ and $\Omega$ is a positive semi-definite matrix that may depend on $\phi$ and/or $\sigma^2$.  When $\Omega$ is the identity matrix, one recovers (\ref{local}).  To fix $\beta_j = 0$ when $\phi_j = 0$,  let $\lambda_j \equiv \phi_j s_j$ for $s_j > 0$, so that when $\phi_j = 0$, the prior variance of $\beta_j$ is set to zero (with prior mean of zero).  \cite{George97}  develops this approach in detail, including the $g$-prior specification, $\Omega(\phi) = \sigma^2(\X_{\phi}^t\X_{\phi})^{-1}$.
 
 Such priors imply that marginally (but not necessarily independently), for $j = 1, \dots, p$,
 \begin{equation}\label{eq:spikeslab2}
\pi(\beta_j \mid \phi_j, \sigma^2, g, s_j) = (1-\phi_j)\delta_0+ \phi_j\N(0, g s_j^2 \omega_j),
\end{equation}
where $\delta_0$ denotes a point mass distribution at zero.  Hierarchical priors of this form are sometimes called ``spike-and-slab" priors ($\delta_0$ is the spike and the continuous full-support  distribution is the slab) or the ``two-groups model" for variable selection. References for this specification include  \cite{mitchell1988bayesian} and \cite{geweke1996variable}, among others.
 
Note that the spike-and-slab approach can be expressed in terms of the prior over $\lambda_j$, by integrating over $\phi$:
 \begin{equation} 
\pi(\lambda_j \mid q) = (1-q) \delta_0 + q P_{\lambda_j},
\end{equation}
where $\mbox{Pr}(\phi_j = 1) = q$, and $P_{\lambda_j}$ is some continuous distribution on $\mathbb{R}^+$.  Of course, $q$ can be given a prior distribution as well; a uniform distribution is common.  This representation transparently embeds model selection priors within the class of local scale mixture of normal distributions.  An important paper exploring the connections between shrinkage priors and model selection priors is \cite{IshwaranRao}, who consider a version of (\ref{eq:spikeslab}) via a specification of $\pi(\lambda_j)$ which is bimodal with one peak at zero and one peak away from zero.  In many respects, this paper anticipated the work of \cite{ParkCasella}, \cite{HansLasso09}, \cite{Horseshoe10}, \cite{Griffin12}, \cite{Polson2012local} and the like.


\subsection{Computational issues in variable selection}
Because posterior sampling is computation-intensive and because variable selection is most  desirable in contexts with many predictor variables, computational considerations are important in motivating and evaluating the approaches above.  The discrete model selection approach and the continuous shrinkage prior approach are both quite challenging in terms of posterior sampling.  

In the model selection setting, for $p > 30$, enumerating all possible models (to compute marginal likelihoods, for example) is beyond the reach of modern capability.  As such, stochastic exploration of the model space is required, with the hope that the unvisited models comprise a vanishingly small fraction of the posterior probability.  \cite{George97} is frank about this limitation; noting that a Markov Chain run of length less than $2^p$ cannot have visited each model even once, they write hopefully that ``it may thus be possible to identify at least some of the high probability values".

\cite{Garcia13} carefully evaluates methods for dealing with this problem and come to compelling conclusions in favor of some methods over others.  Their analysis is beyond the scope of this paper, but we count it as required reading for anyone interested in the variable selection problem in large $p$ settings.  In broad strokes, they find that MCMC approaches based on Gibbs samplers (i.e., \cite{George97}) appear better at estimating posterior quantities---such as the highest probability model, the median probability model, etc---compared to methods based on sampling without replacement (i.e., \cite{hans06sss} and \cite{Clyde11}).

Regarding shrinkage priors, there is no systematic study in the literature suggesting that the above computational problems are alleviated for continuous parameters.  In fact, the results of \cite{Garcia13} (see section 6) suggest that posterior sampling in finite sample spaces is easier than the corresponding problem for continuous parameters, in that convergence to stationarity occurs more rapidly.

Moreover, if one is willing to entertain an extreme prior with $\pi(\phi) = 0$ for $||\phi||_0 > M$ for a given constant $M$, model selection priors offer a tremendous practical benefit:  one never has to invert a matrix larger than $M \times M$, rather than the $p \times p$ dimensional inversions required of a shrinkage prior approach.  Similarly, only vectors up to size $M$ need to be saved in memory and operated upon.  In extremely large problems, with thousands of variables, setting $M = \mathcal{O}(\sqrt{p})$  or $M = \mathcal{O}(\log{p})$ saves considerable computational effort.  For example, this approach is routinely applied to large scale internet data.  Should $M$ be chosen too small, little can be said; if $M$ truly represents one's computational budget, the best model of size $M$ will have to do.

\subsection{Selection: from posteriors to sparsity}
Identifying sparse models (subsets of non-zero coefficients) might be an end in itself, as in the case of trying to isolate scientifically important variables in the context of a controlled experiment.  In this case, a prior with point-mass probabilities at the origin is unavoidable in terms of defining the implicit (multiple) testing problem.  Furthermore, the use of Bayes factors is a well-established methodology for evaluating evidence in the data in favor of various hypotheses.  Indeed, the highest posterior probability model (HPM) is optimal under 0-1 (classification) loss for the selection of each variable. 

If the goal, rather than isolating all and only relevant variables (no matter their absolute size), is to accurately describe the ``important" relationships between predictors and response, then perhaps the model selection route is purely a means to an end.  In this context, a natural question is how to fashion a sparse vector of regression coefficients which parsimoniously characterizes the available data.  \cite{leamer1978specification} is a notable early effort advocating ad-hoc model selection for the purpose of human comprehensibility.  \cite{fouskakis2008comparing},  \cite{DraperBIC}  and \cite{draper2013} represent efforts to define variable importance in real-world terms using subject matter considerations.   A more generic approach is to gauge predictive relevance \citep{gelfand1992model}.  

A widely cited result relating variable selection to predictive accuracy is that of \cite{Barbieri04}.  Consider mean squared prediction error (MSPE), $n^{-1} \E \lbrace \sum_i (\tilde{Y}_i - \tilde{X}_i\hat{\beta})^2 \rbrace$, and recall that the model-specific optimal regression vector is $\hat{\beta}_{\phi} \equiv \E(\beta \mid \mathcal{M}_{\phi}, \mathbf{Y})$.  \cite{Barbieri04}  show that for $\X^t\X$ diagonal, the best predicting model according to MSPE is the model which includes all and only variables with marginal posterior inclusion probabilities greater than 1/2.  This model is referred to as the {\em median probability model} (MPM). Their result holds both for a fixed design $\tilde{\X}$ of prediction points or for stochastic predictors with $\E\lbrace \tilde{X}^t\tilde{X}\rbrace$ diagonal. However, the main condition of their theorem --- $\X^t\X$ diagonal --- is almost never satisfied in practice. Nonetheless, they argue that the median probability model (MPM) tends to outperform the HPM on out-of-sample prediction tasks.   Note that the HPM and MPM are often substantially different models, especially in the case of strong dependence among predictors.  

\cite{George97} suggest an alternative approach, which is to specify a two-point shrinkage prior directly in terms of ``practical significance". Specifically they propose
\begin{equation}
\mbox{Pr}(\lambda_j = s_1) = q; \;\; \mbox{Pr}(\lambda_j = s_2) = (1-q),
\end{equation}
where $s_1$ is a ``large" value reflecting vague prior information about the magnitude of $\beta_j$, and $s_2$ is a ``small" value which biases $\beta_j$ more strongly towards zero.  They suggest setting $s_1$ and $s_2$ such that the prior (mean zero normal) densities are equal at a point $d_j = \Delta Y/\Delta X_j$ where $``\Delta Y$ is the size of an insignificant change in $Y$, and $\Delta X_i$ is the size of the maximum feasible change in $X_j$."  This choice entails that the posterior probability  $\mbox{Pr}(\lambda_j = s_2 \mid \mathbf{Y})$ can be interpreted as the inferred probability that $\beta_j$ is practically significant.  However, this approach does not provide a way to interpret the dependencies that arise in the posterior between the elements of $\lambda_1, \dots, \lambda_p$.

A similar approach, called {\em hard thresholding}, can be employed even if  $\pi(\lambda_j)$ has a continuous density, by stating a classification rule based on posterior samples of $\beta_j$ and $\lambda_j$. For example, \cite{Horseshoe10} suggest setting to zero those coefficients for which $$\E(\kappa_j = 1/(1+\lambda_j^2)\mid \mathbf{Y}) < 1/2.$$    \cite{IshwaranRao} discuss a variety of thresholding rules and relate them to conventional thresholding rules based on ordinary least squares estimates of $\beta$.   As with the approach of \cite{George97}, thresholding approaches do not account for dependencies between the various $\kappa$ variables across predictors, as they are applied marginally.  Indeed, as in \cite{Barbieri04}, the theoretical results of \cite{IshwaranRao} treat only  the orthogonal design case.

\section{Posterior summary variable selection}
None of the priors canvassed above, in themselves, provide sparse model summaries. To go from a posterior distribution to a sparse point estimate requires an additional step, regardless of what prior is used.  Commonly studied approaches tend to neglect posterior dependencies between regression coefficients $\beta_j, j  = 1, \dots, p$ (equivalently, their associated scale factors $\lambda_j$).

In this section we describe a posterior summary based on an expected loss minimization problem.  The loss function is designed to balance prediction ability (in the sense of mean square prediction error) and narrative parsimony (in the sense of sparsity).  The new summary checks three important boxes:

\begin{itemize}
\item it produces sparse vectors of regression coefficients for prediction,
\item it can be applied to a posterior distribution arising from any prior distribution,
\item it explicitly accounts for co-linearity in the matrix of prediction points and dependencies in the posterior distribution of $\beta$.
\end{itemize}
\subsection{The cost of measuring irrelevant variables}
Suppose that collecting information on individual covariates incurs some cost; thus the goal is to make an accurate enough prediction subject to a penalty for acquiring predictively irrelevant facts.

Consider the problem of predicting an $n$-vector of future observables $\tilde{Y} \sim \N(\tilde{\X}\beta, \sigma^2\I)$ at a pre-specified set of design points $\tilde{\X}$.  Assume that a posterior distribution over the model parameters $(\beta$, $\sigma^2)$ has been obtained via Bayesian conditioning, given past data $\mathbf{Y}$ and design matrix $\X$;  denote the density of this posterior by $\pi(\beta, \sigma^2 \mid \mathbf{Y})$.  

It is crucial to note that $\tilde{\X}$ and $\X$ need not be the same.  That is, the locations in predictor space where one wants to predict need not be the same points at which one has already observed past data.  For notational simplicity, we will write $\X$ instead of $\tilde{\X}$ in what follows.  Of course, taking $\tilde{\X} = \X$ is a conventional choice, but distinguishing between the two becomes important in certain cases such as when $p > n$.

Define an optimal action as one which minimizes expected loss $\E(\mathcal{L}(\tilde{Y},\gamma))$, where the expectation is taken over the predictive distribution of unobserved values:
\begin{equation}
f(\tilde{Y}) = \int f(\tilde{Y} \mid \beta, \sigma^2) \pi(\beta, \sigma^2 \mid \mathbf{Y}) d(\beta, \sigma^2).
\end{equation}
As a widely applicable loss function, consider
\begin{equation} \label{Loss}
\mathcal{L}(\tilde{Y},\gamma) = \lambda ||\gamma||_0 + n^{-1}|| \X\gamma - \tilde{Y}||^2_2,
\end{equation}
where $|| \cdot ||_0 = \sum_j \mathbb{1}(\gamma_j \neq 0)$. This loss sums two components, one of which is a ``parsimony penalty" on the action $\gamma$ and  the other of which is the squared prediction loss of the linear predictor defined by $\gamma$. The scalar utility parameter $\lambda$ dictates  how severely we penalize each of these two components, relatively. Integrating over $\tilde{Y}$ conditional on $(\beta, \sigma^2)$ (and overloading the notation of $\mathcal{L}$) gives
\begin{equation}
\label{first}
 \mathcal{L}(\beta,\sigma,\gamma) \equiv \E(\mathcal{L}(\tilde{Y},\gamma))  = \lambda ||\gamma||_0 + n^{-1}|| \X\gamma - \X\beta||^2_2 + \sigma^2.
\end{equation}
Because $(\beta, \sigma^2)$ are unknown, an additional integration over $\pi(\beta, \sigma^2 \mid \mathbf{Y})$ yields 
\begin{equation}
\label{dss0}
 \mathcal{L}(\gamma)  \equiv \E(\mathcal{L}(\beta, \sigma, \gamma)) = \lambda ||\gamma||_0 + \bar{\sigma}^2 +  n^{-1}\mbox{tr}(\X^t\X\Sigma_{\beta}) + n^{-1}||\X\bar{\beta} - \X\gamma||^2_2,
\end{equation}
where $\bar{\sigma}^2 = \E(\sigma^2)$, $\bar{\beta} = \E(\beta)$ and $\Sigma_{\beta} = \mbox{Cov}(\beta)$.  

Dropping constant terms, one arrives at the ``decoupled shrinkage and selection" (DSS) loss function:
\begin{equation}\label{dss}
\mathcal{L}(\gamma) = \lambda ||\gamma||_0 + n^{-1}||\X\bar{\beta} - \X\gamma||^2_2.
\end{equation}

Optimization of the DSS loss function is a combinatorial programming problem depending on the posterior distribution via the posterior mean of $\beta$.  The optimal solution of (\ref{dss}) therefore represents a ``sparsification" of $\bar{\beta}$, which is the theoretically optimal action under pure squared prediction loss.  In this sense, the DSS loss function explicitly trades off the number of variables in the linear predictor with its resulting predictive performance.  Denote this optimal solution by 
 \begin{equation}\label{betalam}
 \beta_{\lambda} \equiv   \mbox{arg min}_{\gamma} \;\;  \lambda ||\gamma||_0  + n^{-1}||\X\bar{\beta} - \X\gamma||^2_2 .
 \end{equation}
 
 Note that the above derivation applies straightforwardly to the selection prior setting via expression (\ref{modelav}) or (equivalently) via the hierarchical formulation in  (\ref{eq:spikeslab2}), which guarantee that $\bar{\beta}$ is well defined marginally across different models.
 


 \subsection{Analogy with high posterior density regions}
Although orthographically (\ref{dss}) resembles expressions used in penalized likelihood methods, the better analogy is a Bayesian high posterior density (HPD) region. Like HPD regions, a DSS summary satisfies a ``comprehensibility criterion"; an HPD interval gives the shortest {\em contiguous} interval encompassing some fixed fraction of the posterior mass, while the DSS summary produces the {\em sparsest} linear predictor which still has reasonable prediction performance.  Like HPD regions, DSS summaries are well defined under any given prior.

To amplify, the DSS optimization problem is well-defined for any {\em posterior} as long as $\bar{\beta}$ exists. Different priors may lead to very different posteriors, potentially with very different means. However, regardless of the precise nature of the posterior (e.g., the presence of multimodality), $\bar{\beta}$ is the optimal summary under squared-error prediction loss, which entails that expression (\ref{betalam}) represents the sparsified solution to the optimization problem given in (\ref{Loss}). 

An important implication of this analogy is the realization that a DSS summary can be produced for a prior distribution directly, in the same way that a prior distribution has a high posterior density region.  The DSS summary requires the user to specify a matrix of prediction points $\tilde{\X}$, but conditional on this choice one can extract sparse linear predictors directly from a prior distribution.

In Section \ref{DSSPlots}, we discuss strategies for using additional features of the posterior $\pi(\beta, \sigma^2 \mid \mathbf{Y})$ to guide the choice of picking $\lambda$.

 \subsection{Computing and approximating $\beta_{\lambda}$}
The counting penalty $||\gamma||_0$ yields an intractable optimization problem for even tens of variables  ($p \approx 30$). This problem has been addressed in recent years by approximating the counting norm with modifications of the $\ell_1$ norm, $||\gamma||_1 = \sum_h |\gamma_h|$, leading to a surrogate loss function which is convex and readily minimized by a variety of software packages.  Crucially, such approximations still yield a sequence of sparse actions (the solution path as a function of $\lambda$), simplifying the $2^p$ dimensional selection problem to a choice between at most $p$ alternatives. The goodness of these approximations is a natural and relevant concern.  Note, however, that the goodness of approximation is entirely non-statistical---the statistical side of the problem has been separately addressed in the formation of the posterior distribution.  This is what is meant by ``decoupled" shrinkage and selection. 

More specifically, recall that DSS requires the evaluation of the optimal solution 
 \begin{equation}
 \beta_{\lambda} \equiv   \mbox{arg min}_{\gamma} \;\;  \lambda ||\gamma||_0  + n^{-1}||\X\bar{\beta} - \X\gamma||^2_2.
 \end{equation}
The most simplistic and yet widely-used approximation is to substitute the $\ell_0$ by the $\ell_1$ norm, which leads to a convex optimization problem for which many implementations are available, in particular the {\tt lars} algorithm (\cite{LARS}). Using this approximation, $\beta_{\lambda}$ can be obtained simply by running the {\tt lars} algorithm using $\bar{\mathbf{Y}} = \X\bar{\beta}$ as the ``data". See also \cite{Bondelletal2012} who similarly use an $\ell_1$ penalty to define a sparse  posterior point estimator.

It is well-known that the $\ell_1$ approximation may unduly ``shrink'' all elements of $\beta_{\lambda}$ beyond the shrinkage arising naturally from the prior over $\beta$. To avoid this potential ``double-shrinkage'' it is possible to explicitly adjust the $\ell_1$ approach towards the desired $\ell_0$ target. Specifically, the local linear approximation argument of \cite{ZouLi08} and \cite{Lv09} advises to solve a surrogate optimization problem (for any $w_j$ near the corresponding $\ell_0$ solution)
 \begin{equation}\label{adapt}
 \beta_{\lambda} \equiv   \mbox{arg min}_{\gamma} \;\;  \sum_j{\frac{\lambda}{|w_j|} |\gamma_j|}  + n^{-1}||\X\bar{\beta} - \X\gamma||^2_2.
 \end{equation}
This approach yields a procedure analogous to the adaptive lasso of \cite{Zhou06} with $\bar{\mathbf{Y}} =\X\bar{\beta}$ in place of $\mathbf{Y}$. In what follows, we use $w_j = \bar{\beta_j}$ (whereas the adaptive lasso uses the least-squares estimate $\hat{\beta}_j$).   The {\tt lars} package in {\tt R} can then be used to obtain solutions to this objective function by a straightforward rescaling of the design matrix. 

In our experience, this approximation successfully avoids double-shrinkage. In fact, as illustrated in the U.S. crime example below, this approach is able to un-shrink coefficients depending on which variables are selected into the model.



For a fixed value of $\lambda$, expression (\ref{dss}) uniquely determines a sparse vector $\beta_{\lambda}$ as its corresponding Bayes estimator. However, choosing $\lambda$ to define this estimator is a non-trivial decision in its own right.  Section \ref{DSSPlots} considers how to use the posterior distribution $\pi(\beta, \sigma^2 \mid \mathbf{Y})$ to illuminate the trade-offs implicit in the selection of a given value of $\lambda$.

 \section{Selection summary plots}\label{DSSPlots}
How should one think about the generalization error across possible values of $\lambda$? Consider first two extreme cases. When $\lambda = 0$, the solution to the DSS optimization problem is simply the posterior mean:  $\beta_{\lambda = 0} \equiv \bar{\beta}$.  Conversely, for very large $\lambda$, the optimal solution will be the zero vector, $\beta_{\lambda = \infty} = 0$, which will have expected prediction loss equal to the marginal variance of the response $Y$ (which will depend on the predictor points in question).   Letting $\lambda$ depend on sample size so that $\lambda_n \rightarrow 0$ faster than the posterior distribution concentrates about the true parameter value, will give a consistent estimator \citep{Bondelletal2012}. But in applied scenarios with finite samples, the sparsity of $\beta_{\lambda}$ depends directly on the choice of $\lambda$, making its selection an important consideration.  
 
A sensible way to judge the goodness of $\beta_{\lambda}$ in terms of prediction is relative to the predictive performance of $\beta$---were it known---which is the optimal linear predictor under squared-error loss. The relevant scale for this comparison is dictated by $\sigma^2$, which quantifies the best one can hope to do even if $\beta$ were known.  With these benchmarks in mind, one wants to address the question: how much predictive deterioration is a result of sparsification?
 
The remainder of this section defines three plots that can be used by a data analyst to visualize the predictive deterioration across various values of $\lambda$. The first plot concerns a measure of ``variation explained", the second plot considers the excess prediction loss on the scale of the response variable, and the final plot looks at the magnitude of the elements of $\beta_{\lambda}$. Throughout, we will preprocess the outcome variable and covariates to be centered at zero and scaled to unit variance.    
 
 \subsection{Variation explained of a sparsified linear predictor}
 Define the ``variation-explained" at design points $\X$ (perhaps different than those seen in the data sample used to form the posterior distribution) as: 
 \begin{equation}
 \rho^2 = \frac{n^{-1}||\X\beta||^2}{n^{-1}||\X\beta||^2 + \sigma^2}.
 \end{equation}
Denote by
 \begin{equation}\label{varex}
 \rho_{\lambda}^2 = \frac{n^{-1}||\X\beta||^2}{n^{-1}||\X\beta||^2 + \sigma^2 + n^{-1}||\X\beta - \X\beta_{\lambda}||^2}
 \end{equation}
the analogous quantity for the sparsified linear predictor $\beta_{\lambda}$.   The gap between $\beta_{\lambda}$ and $\beta$ due to sparsification is tallied as a contribution to the noise term, which decreases the variation explained. This quantity has the benefit of being directly comparable to the ubiquitous $R^2$ metric of model fit familiar to users of statistical software and least-squares theory.  

Posterior samples of $\rho^2_{\lambda}$ can be obtained as follows.  First, solve (\ref{adapt}) by applying the {\tt lars} algorithm with inputs $w_j = \bar{\beta_j}$ and $\mathbf{Y} = \X\bar{\beta}$.  A single run of this algorithm will produce a sequence of solutions $\beta_{\lambda}$ for a range of $\lambda$ values. (Obtaining draws of $\rho_{\lambda}^2$ using a model selection prior requires posterior samples from $(\beta, \sigma^2)$ marginally across models.) Second, for each element in the sequence of $\beta_{\lambda}$'s, convert posterior samples of $(\beta, \sigma^2)$ into  samples of $\rho_{\lambda}^2$ via definition (\ref{varex}).   Finally, plot the expected value and 90\% credible intervals of $\rho^2_{\lambda}$ against the model size, $||\beta_{\lambda}||_\lambda$.  The posterior mean of $\rho^2_0$ may be overlaid as a horizontal line for benchmarking purposes; note that even for $\lambda=0$ (so that $\beta_{\lambda=0} = \bar{\beta}$), the corresponding variation explained, $\rho^2_0$, will have a (non-degenerate) posterior distribution induced by the posterior distribution over  $(\beta, \sigma^2)$.  

Variation explained sparsity summary plots depict the posterior uncertainty of $\rho_{\lambda}^2$, thus providing a measure of confidence concerning the predictive goodness of the sparsified vector.  In these plots, one often observes that the sparsified variation explained does not deteriorate ``statistically significantly" in the sense that the credible interval for $\rho_{\lambda}^2$ overlaps the posterior mean of the unsparsified variation explained. 


\subsection{Excess error of a sparsified linear predictor}
Define the ``excess error" of a sparsified linear predictor $\beta_{\lambda}$ as
\begin{equation}\label{excess}
\psi_{\lambda} = \sqrt{n^{-1}||\X\beta_{\lambda} - \X\beta||^2 + \sigma^2} - \sigma.
\end{equation}
This metric of model fit, while less widely used than variation explained, has the virtue of being on the same scale as the response variable. Note that excess error attains a minimum of zero precisely when $\beta_{\lambda} = \beta$.  As with the variation explained, the excess error is a random variable and so has a posterior distribution.  By plotting the mean and 90\% credible intervals of the excess error against model size (corresponding to increasing values of $\lambda$), one can see at a glance the degree of predictive deterioration incurred by sparsification.  Samples of $\psi_{\lambda}$ can be obtained analogously to the procedure for producing samples of $\rho^2_{\lambda}$, but using (\ref{excess}) in place of (\ref{varex}).


\subsection{Coefficient magnitude plot}
In addition to the two previous plots, it is instructive to examine which variables remain in the model at different levels of sparsification, which can be achieved simply by plotting the magnitude of each element of $\beta_{\lambda}$ as $\lambda$ (hence model size) varies.  However, using $\lambda$ or $||\beta_{\lambda}||_0$ for the horizontal axis can obscure the real impact of the sparsification because the predictive impact of sparsification is non-constant. That is, the jump from a model of size 7 to one of size 6, for example, may correspond to a negligible predictive impact, while the jump from model of size 3 to a model of size 2 could correspond to considerable predictive deterioration.  Plotting the magnitude of the elements of $\beta_{\lambda}$ against the corresponding excess error $\psi_{\lambda}$ gives the horizontal axis a more interpretable scale.  

\subsection{A heuristic for reporting a single model}\label{heuristic}
The three plots described above achieve a remarkable consolidation of information hidden within the posterior samples of $\pi(\beta, \sigma^2 \mid \mathbf{Y})$. They relate sparsification of a linear predictor to the associated loss in predictive ability, while keeping the posterior uncertainty in these quantities in clear view.  Nonetheless, in many situations one would like a procedure that yields a {\em single} linear predictor.  

For producing a single-model linear summary, we propose the following heuristic: report the sparsified linear predictor corresponding to the smallest model whose 90\% $\rho^2_{\lambda}$ credible interval contains $\E(\rho_0^2)$.  In words, we want the smallest linear predictor whose predictive ability (practical significance) is not statistically different than the full model's. 

This choice leans on convention, for example, to determine the 90\% level (rather than say the 75\% or 95\%).  However, this is true of alternative methods such as hard thresholding or examination of marginal inclusion probabilities, which both require similar conventional choices to be determined.  The DSS model selection heuristic offer a crucial benefit over these approaches, though---it explicitly includes a design matrix of predictors into its very formulation.  Standard thresholding rules and methods such as the median probability model approach are instead defined on a one-by-one basis, which does not explicitly account for colinearity in the predictor space.  (Recall that both the thresholding rules studied in \cite{IshwaranRao} and the median probability theorems of \cite{Barbieri04} restrict their analysis to the orthogonal design situation.)

In the DSS approach to model selection, dependencies in the predictor space appear both in the formation of the posterior and also in the definition of the loss function.  In this sense, while the response vector $\mathbf{Y}$ is only ``used once" in the formation of the posterior, the design information may be ``used twice", both in defining the posterior and also in defining the loss function.  Note that this is completely kosher in the sense that the model is conditional on $\X$ in the first place. Note also that the DSS loss function may be based on a predictor matrix {\em different} than the one used in the formation of the posterior. 

\subsection*{Example: U.S. crime dataset ($p=15$, $n=47$)}
The U.S. crime data of \cite{Vandaele78} appears in \cite{Raftery97} and \cite{Clyde11} among others. The dataset consists of $n=47$ observations on $p=15$ predictors.  As in earlier analyses we log transform all continuous variables. We produce DSS selection summary plots for three different priors:  {\em(i)} the horseshoe prior, {\em (ii)} the robust prior of \cite{Berger12} with uniform model probabilities, and (iii) a $g$-prior with $g=n$ and model probabilities as suggested in \cite{ScottBerger06}.  With $p=15 < 30$, we are able to evaluate marginal likelihoods for all models under the model selection priors (ii) and (iii).

We use these particular priors not to endorse them, but merely as representative examples of widely-used specifications. 

\begin{figure}[t!]
\begin{center}
\includegraphics[width=6in]{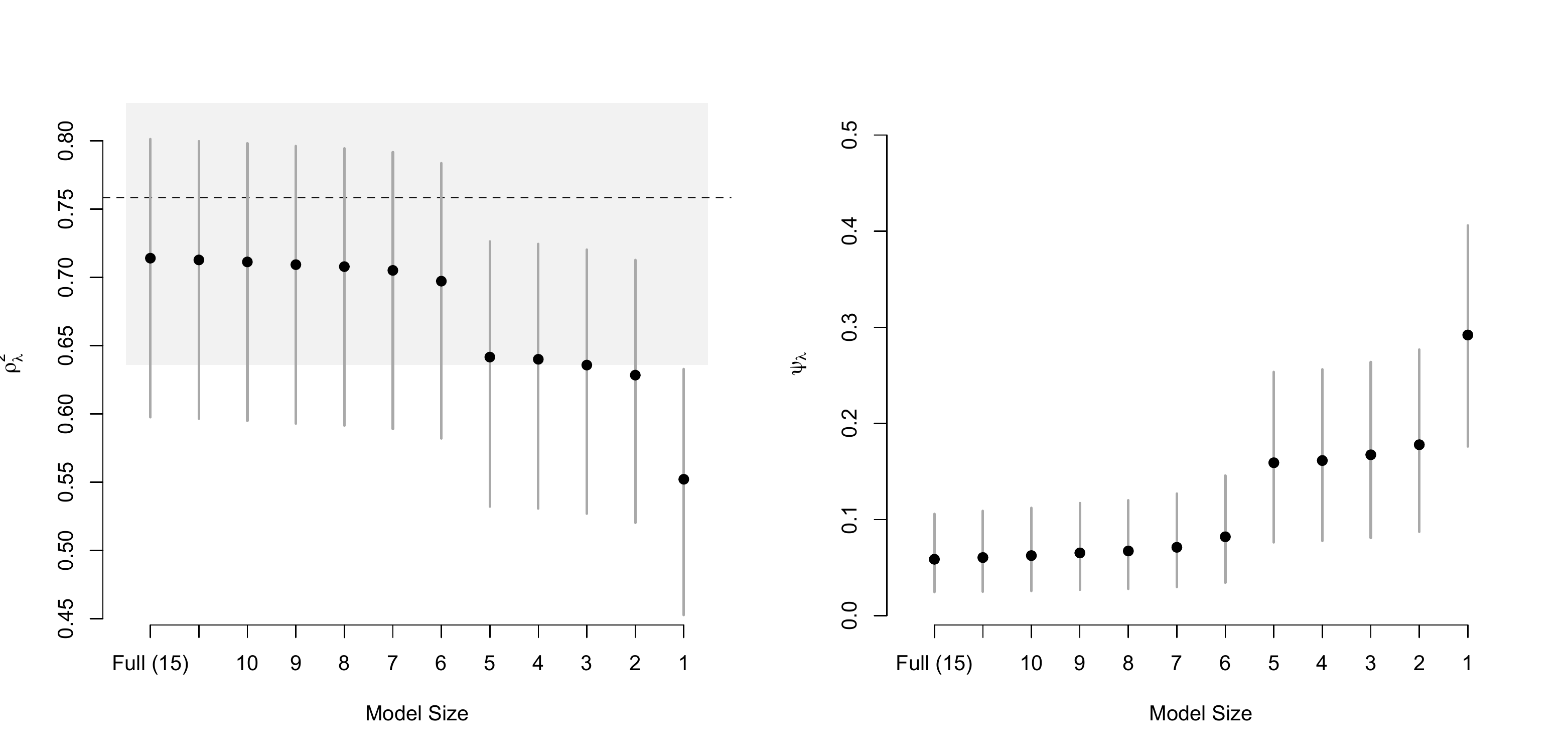}\\
\includegraphics[width=4in]{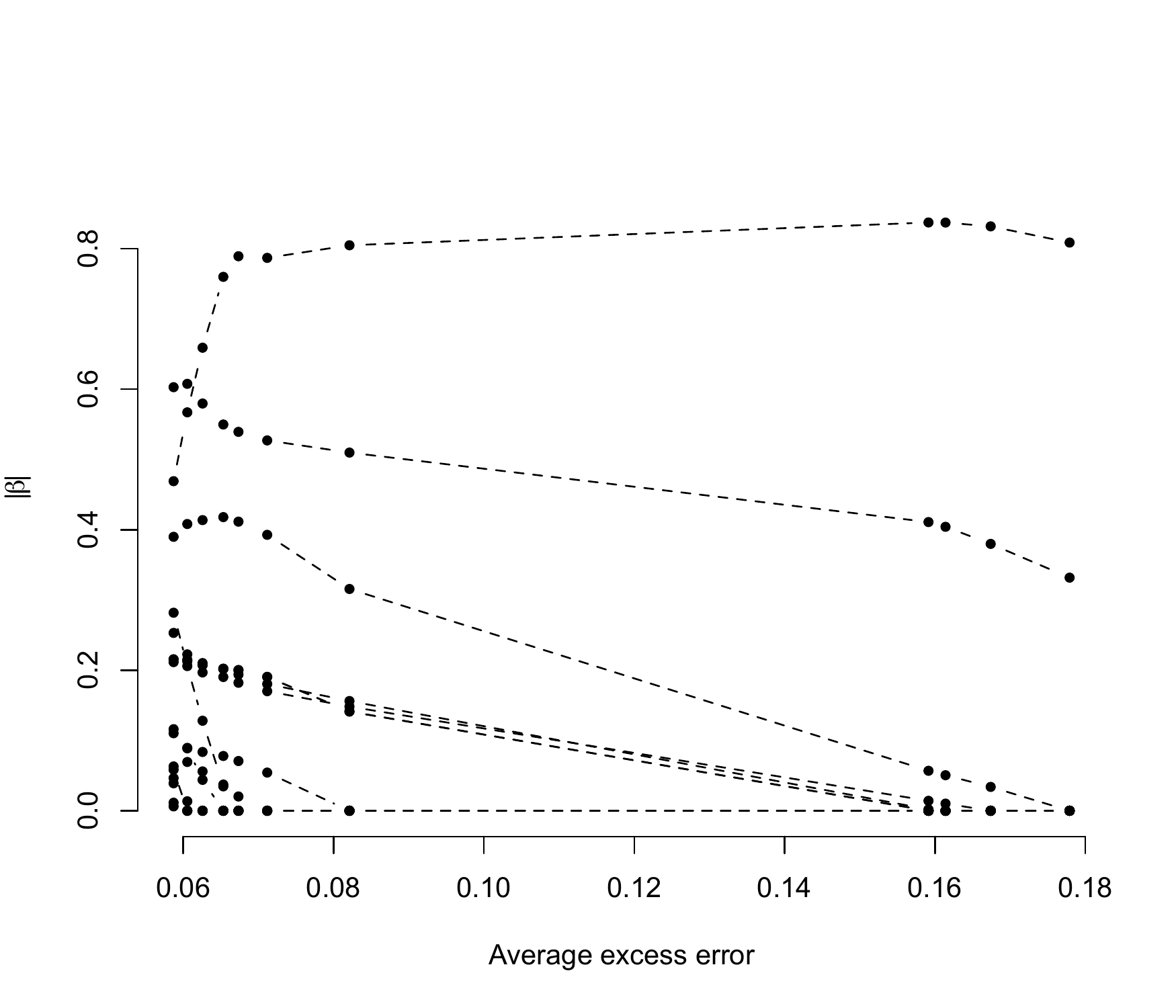}
\caption{U.S. Crime Data: DSS plots under the horseshoe prior.}
\label{uscrime1}
\end{center}
\end{figure}
\begin{figure}[b!]
\begin{center}
\includegraphics[width=6in]{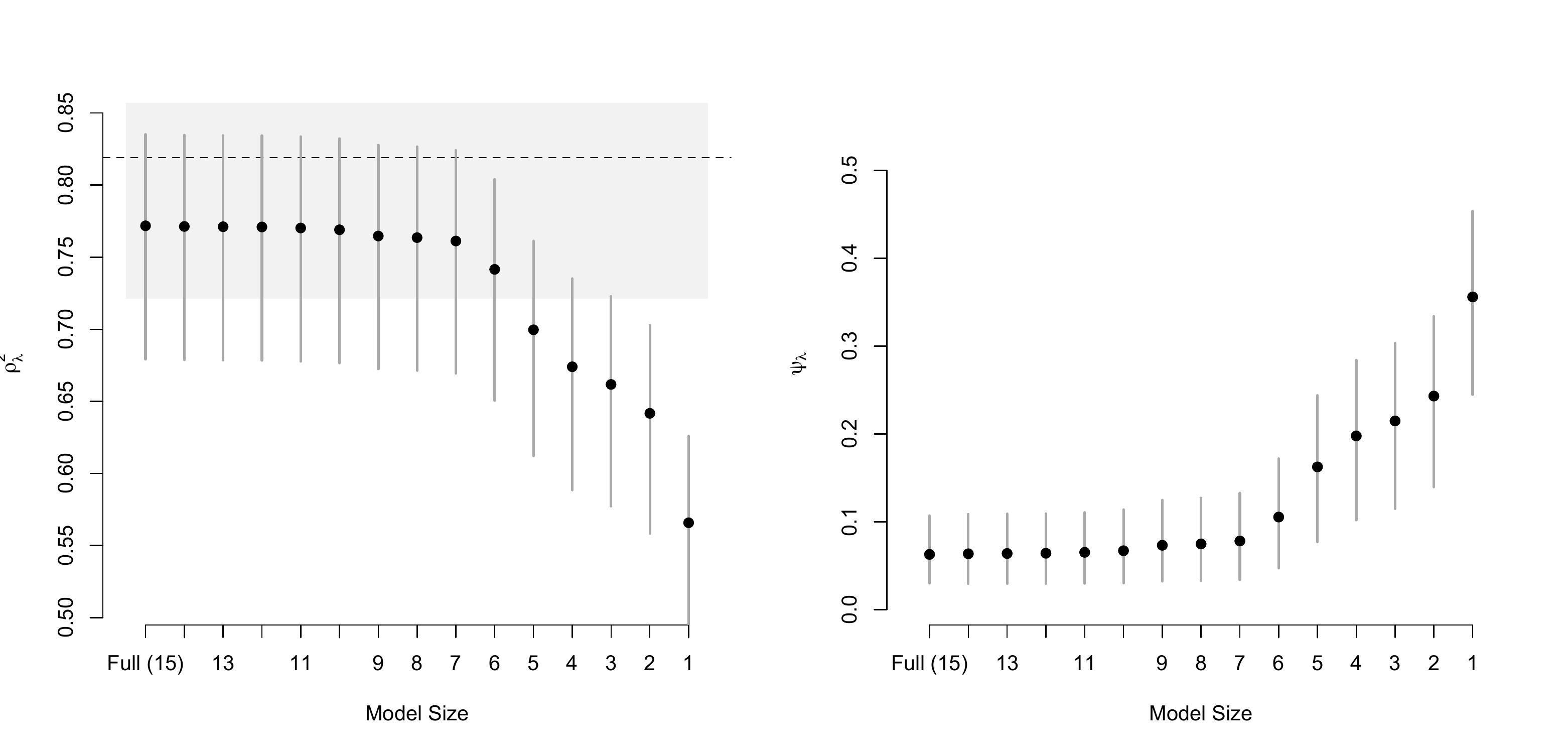}\\
\includegraphics[width=6in]{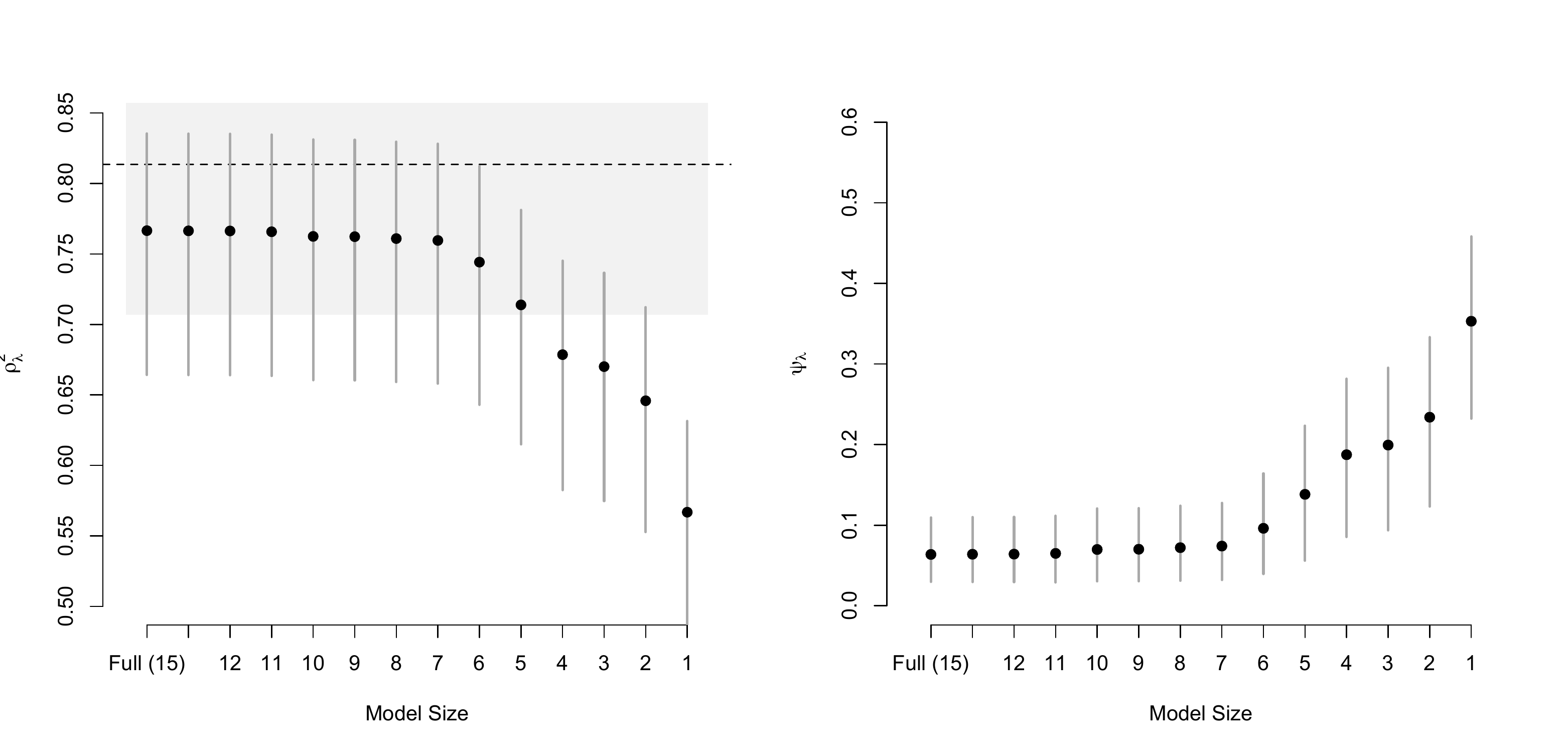}
\caption{U.S. Crime Data: DSS plot under the ``robust'' prior of \cite{Berger12} (top row) and under a $g$-prior with $g=n$ (bottom row). All $2^{15}$ models were evaluated in this example.}
\label{uscrime2}
\end{center}
\end{figure}

Figures \ref{uscrime1} and \ref{uscrime2} show the resulting DSS plots under each prior.  Notice that with this data set the prior choice has an impact; the resulting posteriors for $\rho^2$ are quite different. For example, under the horseshoe prior we observe a significantly larger amount of shrinkage, leading to a posterior for $\rho^2$ that concentrates around smaller values as compared to the results in Figure \ref{uscrime2}. Despite this difference, a conservative reading of the plots would lead to the same conclusion in either situation: the 7-variable model is essentially equivalent (in both suggested metrics, $\rho^2$ and $\psi$) to the full model.  

To use these plots to produce a {\em single} sparse linear predictor for the purpose of data summary, we employ the heuristic described in Section \ref{heuristic}. Table \ref{table:uscrime} compares the resulting summary to the model chosen according to the median probability model criterion. Notably, the DSS heuristic yields the same 7-variable model under all three choices of prior.  In contrast,  the HPM is the full model, while the MPM gives either an 11-variable or a 7-variable model depending on which prior is used.  Both the HPM and MPM under the robust prior choice would include variables with low statistical and practical significance. 

Notice also that the MPM under the robust prior contains four variables with marginal inclusion probabilities near $1/2$. The precise numerics of these quantities is highly prior dependent and sensitive to search methods when enumeration is not possible.  Accordingly, the MPM model in this case is highly unstable.  By focusing on metrics more closely related to practical significance, the DSS heuristic provides more stable selection, returning the same 7-variable model under all prior specifications in this example. As such, this data set provides a clear example of statistical significance---as evaluated by standard posterior quantities---overwhelming practical relevance. The summary provided by a selection summary plot makes an explicit distinction between the two notions of relevance, providing a clear sense of the predictive cost associated with dropping a predictor. 

\begin{figure}[t!]
\begin{center}
\includegraphics[width=5in]{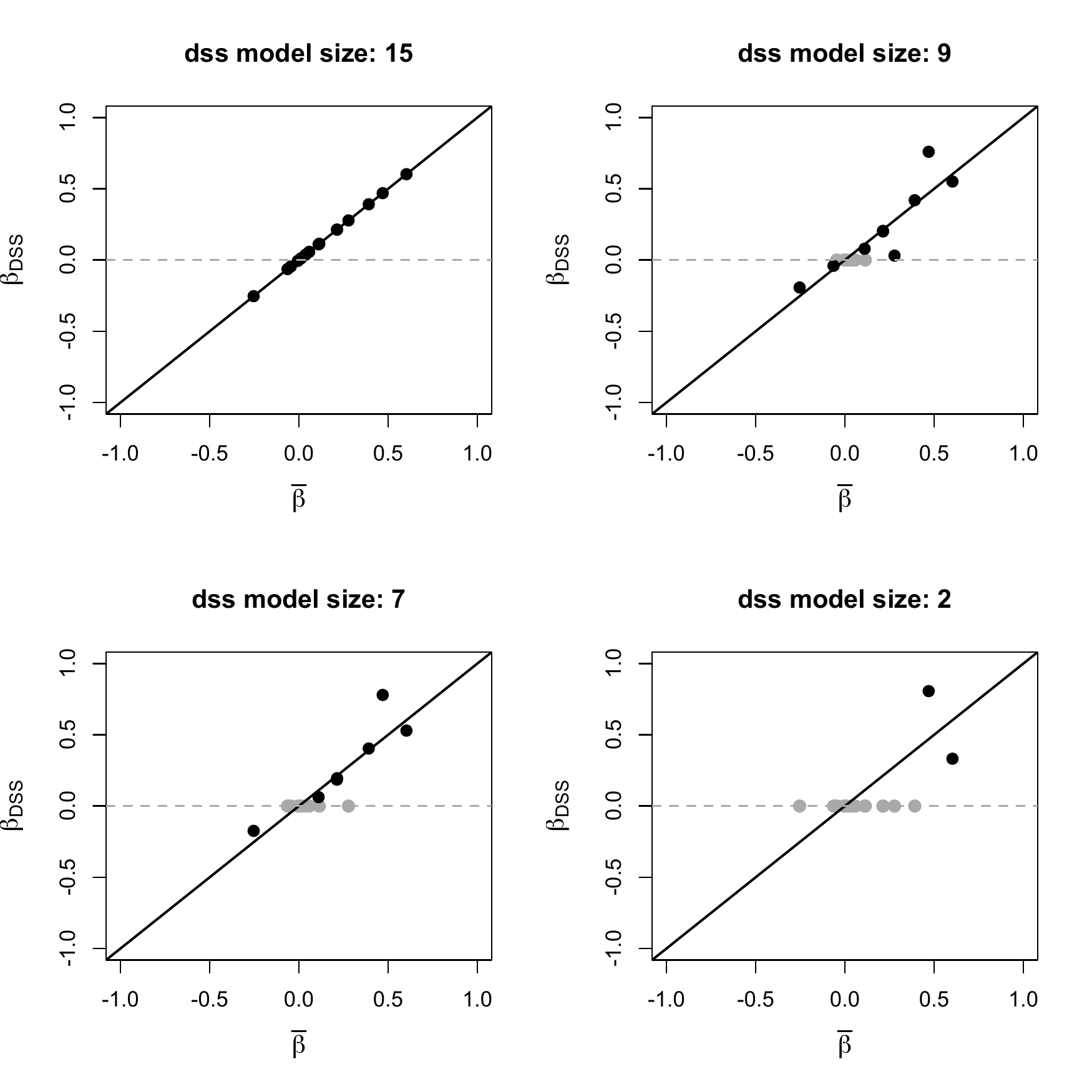}
\caption{U.S. Crime data under the horseshoe prior: $\bar{\beta}$ refers to the posterior mean while $\beta_{DSS}$ is the value of $\beta_{\lambda}$ under different values of $\lambda$ such that different number of variables are selected.}
\label{uscrime3}
\end{center}
\end{figure}

Finally, notice that there is no evidence of ``double-shrinkage".  That is, one might suppose that DSS  penalizes coefficients twice, once in the prior and again in the sparsification process, leading to unwanted attenuation of large signals. However, double-shrinkage would not occur if the $\ell_0$ penalty were being applied exactly, so any unwanted attenuation is attributable to the imprecision of the surrogate optimization in (\ref{adapt}). In practice, we observe that the adaptive lasso-based approximation exhibits minimal evidence of double-shrinkage.  Figure \ref{uscrime3} displays the resulting values of $\beta_{\lambda}$  in the U.S. crime example plotted against the posterior mean (under the horseshoe prior).  Notice that, moving from larger to smaller models, no double-shrinkage is apparent. In fact, we observe re-inflation or ``unshrinkage'' of some coefficients as one progresses to smaller models, as might be expected under the $\ell_0$ norm.

\begin{table}[h!] 
\begin{center}
\begin{tabular}{lccccc}\toprule
&DSS-HS (7) &DSS-Robust (7) & DSS-$g$-prior (7) & MPM (Robust) &MPM ($g$-prior)\\
\midrule
M&$\bullet$ &$\bullet$  &$\bullet$& {\bf0.89}& {\bf0.85} \\

So& -- &-- & -- &  0.39& 0.27 \\

Ed & $\bullet$ & $\bullet$  &$\bullet$&  {\bf 0.97} & {\bf0.96}\\

Po1& $\bullet$& $\bullet$&$\bullet$& {\bf 0.71}& {\bf0.68}\\

Po2 & --& --& --& {\bf 0.52}& 0.45\\

LF& --& --& --&0.36&0.22\\

M.F& -- &--&  --& 0.38&0.24\\

Pop& --&-- &-- &{\bf 0.51}&0.40\\

NW&$\bullet$  &$\bullet$&$\bullet$ &{\bf 0.77}& {\bf0.70}\\

U1&-- & --& --&0.39&0.27\\

U2&$\bullet$ &$\bullet$& $\bullet$&{\bf 0.71}& {\bf0.63}\\

GDP&-- & --& --& {\bf 0.52}& 0.39\\

Ineq&$\bullet$ &$\bullet$& $\bullet$& {\bf 0.99}& {\bf0.99}\\

Prob&$\bullet$ &$\bullet$& $\bullet$& {\bf 0.91}& {\bf0.88}\\

Time&-- & --& --&{\bf 0.52}&0.40\\

\midrule
\addlinespace
$R^2_{MLE}$ & 82.6\% & 82.6\% &82.6\% &  85.4\% &82.6\% \\
\bottomrule
\end{tabular}
\end{center}
\caption{Selected models by different methods in the U.S. crime example. The MPM column displays marginal inclusion probabilities with the numbers in bold associated with the variables included in the median probability model. The $R^2_{mle}$ row reports the traditional in-sample percentage of variation explained of the least-squares fit based on only the variables in a given column.
}
\label{table:uscrime}
\end{table}

\subsection*{Example: Diabetes dataset ($p=10$, $n=447$)}
The diabetes data was used to demonstrate the {\tt lars} algorithm in \cite{LARS}. The data consist of $p =10$ baseline measurements on $n = 442$ diabetic patients; the response variable is a numerical measurement of disease progression.  As in \cite{LARS}, we work with centered and scaled predictor and response variables. In this example we only used the robust prior of \cite{Berger12}. The goal is to focus on the sequence in which the variables are included and to illustrate how DSS provides an attractive alternative to the median probability model.  

Table \ref{table:diabetes} shows the variables included in each model in the DSS path up to the 5-variable model. The DSS plots in this example (omitted here) suggest that this should be the largest model under consideration. The table also reports the median probability model.

Notice that marginal inclusion probabilities do not necessarily offer a good alternative to rank variable importance, particularly in cases where the predictors are highly colinear.  This is evident in the current example in the ``dilution" of inclusion probabilities of the variables with the strongest dependencies in this dataset: TC, LDL, HDL, TCH and LTG. It is possible to see the same effect in the rank of high probability models, as most models on the top of the list represent distinct combinations of correlated predictors. In the sequence of models from DSS, variables HDL and LTG are chosen as the representatives for this group. 

Meanwhile, a variable such as Sex (highly correlated with BMI) appears with marginal inclusion probability of 0.98, and yet its removal from DSS (five-variable) leads to only a minor increase in the model's predictive ability. Thus the diabetes data offer a clear example where statistical significance can overwhelm practical relevance if one looks only at standard Bayesian outputs. The summary provided by DSS makes a distinction between the two notions of relevance, providing a clear sense of the predictive cost associated with dropping a predictor.\\

\begin{table}[h!] 
\begin{center}
\begin{tabular}{lcccccc}\toprule
&DSS-Robust (5) &DSS-Robust  (4) & DSS-Robust  (3)&DSS-Robust  (2)&DSS-Robust (1)& MPM ({\tt Robust})\\
\midrule
Age&-- &--  & --&-- &--& 0.08 \\

Sex& $\bullet$&-- &--& --&--& {\bf 0.98} \\

BMI &$\bullet$ &$\bullet$ &$\bullet$&$\bullet$&$\bullet$&{\bf 0.99}\\

MAP& $\bullet$& $\bullet$&$\bullet$& --&--&{\bf 0.99}\\

TC & --& --& --&--&--& {\bf 0.66}\\

LDL& --& --& --&--&--& 0.46\\

HDL & $\bullet$ &$\bullet$&  --& --&--&{\bf 0.51}\\

TCH& --&-- &-- &--  &--&0.26\\

LTG&$\bullet$  &$\bullet$&$\bullet$  &$\bullet$&--&{\bf 0.99}\\

GLU&-- & --& --& -- &--&0.13\\

\midrule
\addlinespace
$R^2_{MLE}$ & 50.8\% & 49.2\% &48.0\% & 45.9\%&34.4\%&51.3\% \\
\bottomrule
\end{tabular}
\end{center}
\caption{Selected models by DSS and model selection prior in the Diabetes example. The MPM column displays marginal inclusion probabilities, and the numbers in bold are associated with the variables included in the median probability model. 
}
\label{table:diabetes}
\end{table}

\subsection*{Example: protein activation dataset ($p=88$, $n=96$)}
The protein activity dataset is from \cite{Clyde11}.  This example differs from the previous example in that with $p=88$ predictors, the model space can no longer be exhaustively enumerated. In addition, correlation between the potential predictors is as high as 0.99, with 17 pairs of variables having correlations above 0.95.  For this example, the horseshoe prior and the robust prior are considered. To search the model space, we use a conventional Gibbs sampling strategy as in \cite{Garcia13} (Appendix A), based on \cite{George97}. 

\begin{figure}[t!]
\begin{center}
\includegraphics[width=7in]{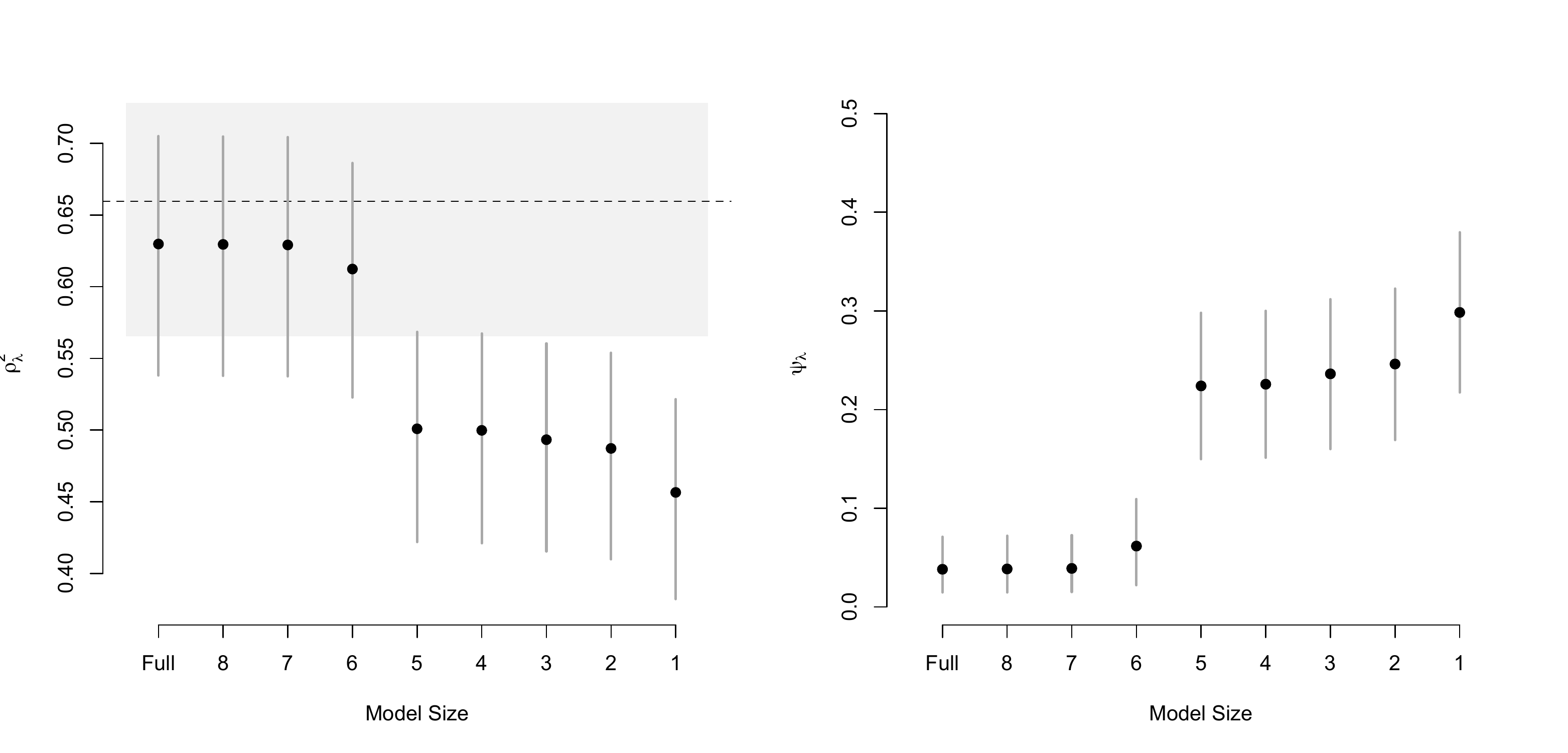}\\
\includegraphics[width=7in]{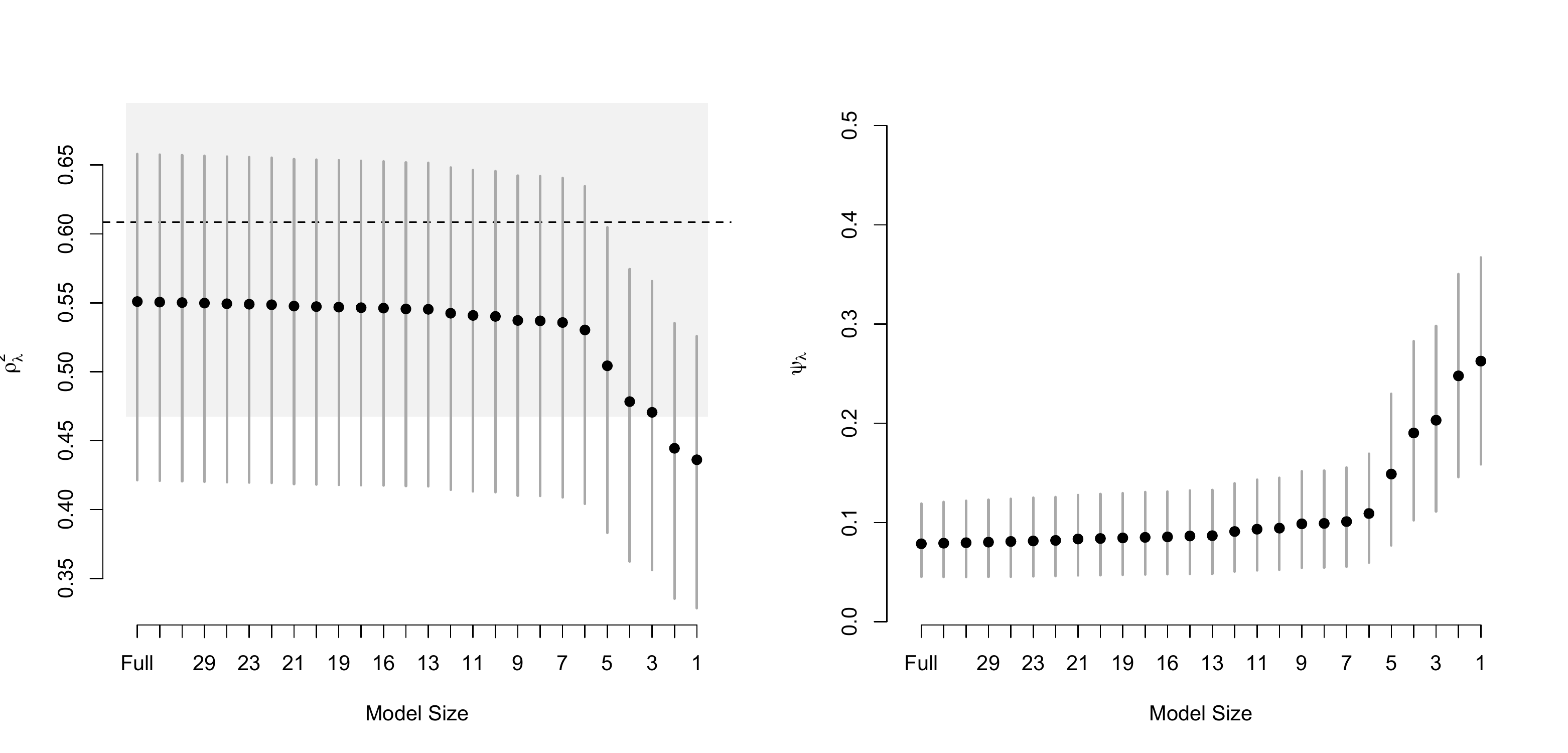}
\caption{Protein Activation Data: DSS plots under model selection priors (top row) and under shrinkage priors (bottom row).}
\label{protein1}
\end{center}
\end{figure}

Figure \ref{protein1} shows the DSS plots under the two priors considered. Once again, the horseshoe prior leads to smaller estimates of $\rho^2$. And once again, despite this difference, the DSS heuristic returns the same (7) predictors under both priors.  On this data set, the MPM under the Gibbs search (as well as the HPM and MPM given by {\tt BAS}) coincide with the DSS summary model.

\subsection*{Example: protein activation dataset ($p=88$, $n=80$)}
\begin{figure}[h!]
\begin{center}
\includegraphics[width=7in]{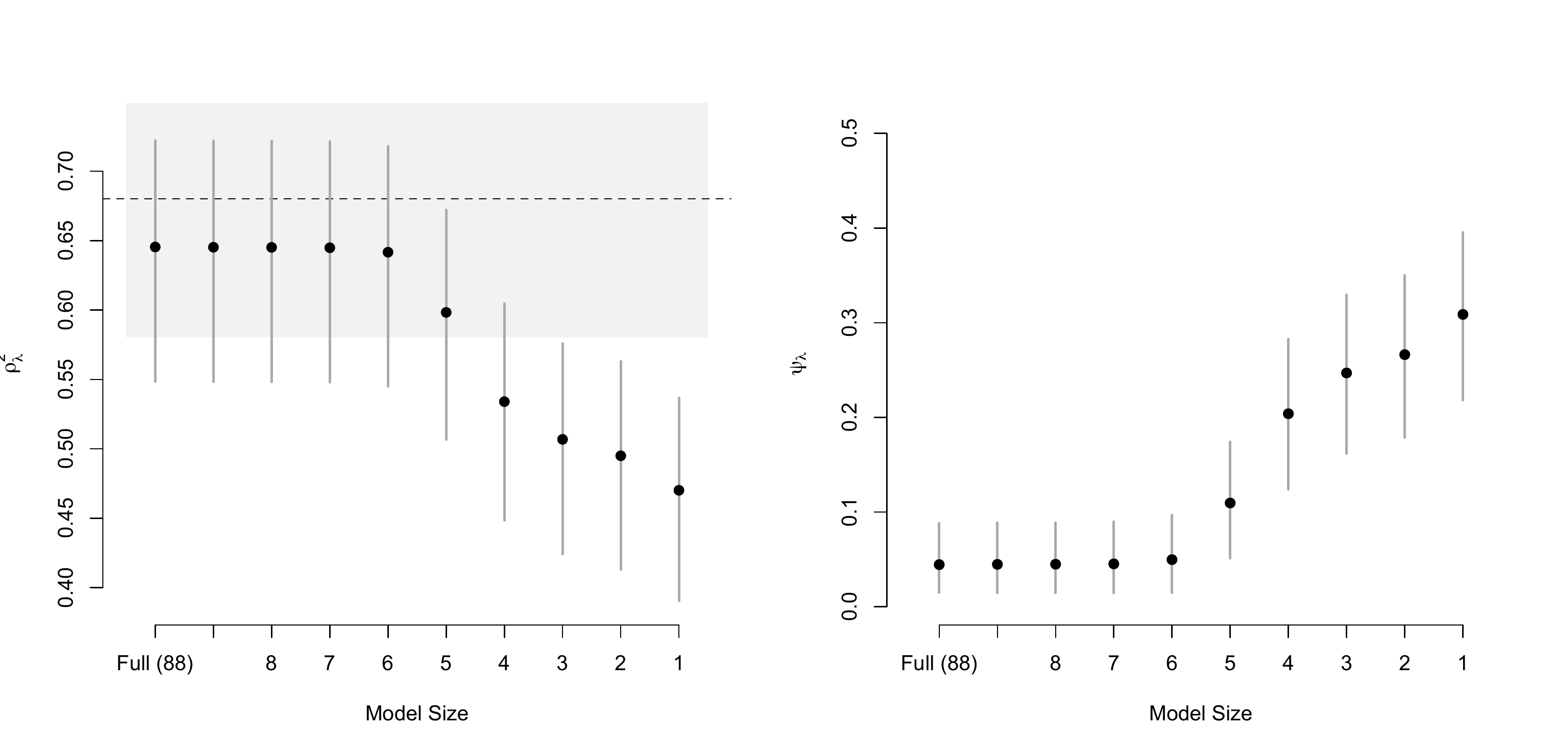}\\
\caption{Protein Activation Data ($p>n$ case): DSS plots under model selection priors}
\label{protein2}
\end{center}
\end{figure}

 To explore the behavior of DSS in the $p > n$ regime, we modify the previous example by randomly selecting a subset of $n=80$ observations from the original dataset.  These 80 observations are used to form our posterior distribution.  To define the DSS summary, we take $\tilde{\X}$ to be the entire set of 96 predictor values.  For simplicity we only use the robust model selection prior. Figure \ref{protein2} shows the results; with fewer observations, smaller models don't give up as much in the $\rho^2$ and $\psi$ scales as the original example. A conservative read of the DSS plots leads to the same 6-variable model, however, in this limited information situation, the models with 5 or 4 variables are competitive.
 One important aspect of Figure \ref{protein2} is that even working in the $p>n$ regime, DSS is able to evaluate the performance and provide a summary of models of any dimension up to the full model. This is accomplished even in this situations where by using the robust prior, the posterior was limited to models up to dimension $n-1$. In order for this to be achieved all DSS needs is the number of points in $\mathbf{X}$ to be larger than $p$. In situations where not enough points are available in the dataset, all the user needs to do is to add (arbitrary and without loss of generality) representative points in which to make predictions about potential $Y$.  
  
\section{Discussion}

A detailed examination of the previous literature reveals that sparsity can play many roles in a statistical analysis---model selection, strong regularization, and improved computation, for example.  A central, but often implicit, virtue of sparsity is that human beings find fewer variables easier to think about.  

When one desires sparse model summaries for improved comprehensibility, prior distributions are an unnatural vehicle for furnishing this bias.  Instead, we describe how to use a decision theoretic approach to induce sparse posterior model summaries. Our new loss function resembles the popular penalized likelihood objective function of the lasso estimator, but its interpretation is very different.  Instead of a regularizing tool for estimation, our loss function is a posterior summarizer with an explicit parsimony penalty.  To our knowledge this is the first such loss function to be proposed in this capacity.  Conceptually, its nearest forerunner would be high posterior density regions, which summarize a posterior density while satisfying a compactness constraint.

Unlike hard thresholding rules, our selection summary plots convey posterior uncertainty associated with the provided sparse summaries.  In particular, posterior correlation between the elements of $\beta$ impacts the posterior distribution of the sparsity degradation metrics $\rho^2$ and $\psi$.  While the DSS approach does not ``automate" the problem of determining $\lambda$ (and hence $\beta_{\lambda}$), they do manage to distill the posterior distribution into a graphical summary that reflects the posterior uncertainty in the predictive degradation due to sparsification.  Furthermore, they explicitly integrate information about the possibly non-orthogonal design space in ways that standard thresholding rules and marginal probabilities do not.

As a summary device, these plots can be used in conjunction with whichever prior distribution is most appropriate to the applied problem under consideration.  As such, they complement recent advances in Bayesian variable selection and shrinkage estimation and will benefit from future advances in these areas. 

We demonstrate how to apply the summary selection concept to logistic regression and Gaussian graphical models in a brief appendix.

\newpage

\bibliographystyle{abbrvnat}
\bibliography{DSS}
\appendix
\section{Extensions}
\subsection{Selection summary in logistic regression}
Selection summary can be applied outside the realm of normal linear models as well. This section explicitly shows how to  extend the approach to logistic regression and provides an illustration on real data. 

Although one has many choices for judging predictive accuracy, it is convenient to note that squared prediction loss is precisely the negative log likelihood in the normal linear model setting, which suggests the following generalization of (\ref{Loss}):
\begin{equation}
\mathcal{L}(\tilde{Y}, \gamma) = \lambda ||\gamma||_0 - n^{-1} \log\left[ f(\tilde{Y}, \X, \gamma) \right]
\end{equation}
where $f(\tilde{Y}, \gamma)$ denotes the likelihood of $\tilde{Y}$ with ``parameters" $\gamma$.  

In the case of a binary outcome vector using a logistic link function, the generalized DSS loss becomes
  \begin{equation}\label{logisticDSS}
\mathcal{L}(\tilde{Y}, \gamma) =\lambda ||\gamma||_0 + n^{-1}\sum_{i=1}^n \left(\tilde{Y}_i \X_i \gamma - \log{(1 + \exp{(\X_i\gamma))}}\right)  .
\end{equation}
Taking expectations yields   
  \begin{equation}\label{logisticDSS}
\mathcal{L}(\tilde{Y}, \bar{\pi}) =\lambda ||\gamma||_0 + n^{-1}\sum_{i=1}^n \left(\bar{\pi_i}\X_i \gamma - \log{(1 + \exp{(\X_i\gamma))}}\right),
\end{equation}
where $\bar{\pi}_i$ is the posterior mean probability that $\tilde{Y}_i = 1$.  To help interpret this formula, note that it can be rewritten as a weighted logistic regression as follows.  For each observed $X_i$, associate a pair of pseudo-responses $Z_i = 1$ and $Z_{i+n} = 0$ with weights $w_i = \bar{\pi}_i$ and $w_{i+n} = 1- \bar{\pi}_i$ respectively.  Then $\bar{\pi}_i  \mbox{X}_i\gamma - \log{(1 + \exp{(\mbox{X}_i\gamma))}}$ may be written as
\begin{equation}
 \Big [ w_i Z_i \mbox{X}_i\gamma - w_i\log{(1 + \exp{(\mbox{X}_i\gamma))}}\Big] +  \Big[w_{i+n} Z_{i+n} \mbox{X}_i\gamma - w_{i+n}\log{(1 + \exp{(\mbox{X}_i\gamma))}}\Big].
 \end{equation}
Thus, optimizing the DSS logistic regression loss is equivalent to finding the penalized maximum likelihood of a weighted logistic regression where each point in predictor space has a response $Z_i = 1$, given weight $\bar{\pi}_i$, and a counterpart response $Z_i = 0$, given weight $1 - \bar{\pi}_i$.  The observed data determines $\bar{\pi}_i$ via the posterior distribution.  As before, if we replace (\ref{logisticDSS}) by the surrogate $\ell_1$ norm 
  \begin{equation}\label{logisticDSS1}
\mathcal{L}(\tilde{Y}, \bar{\pi}) =\lambda ||\gamma||_1 + n^{-1}\sum_{i=1}^n \left(\bar{\pi_i}\X_i \gamma - \log{(1 + \exp{(\X_i\gamma))}}\right),
\end{equation}
then an optimal solution can be computed via the {\tt R} package GLMNet (\cite{GLMNet}). 

The DSS summary selection plot may be adapted to logistic regression by defining the excess error as
\begin{equation}
\psi_{\lambda} = \sqrt{n^{-1} \sum_i \pi_i - 2\pi_{\lambda,i}\pi_i + \pi_{\lambda,i}^2} - \sqrt{n^{-1} \sum_i \pi_i(1 - \pi_i)}
\end{equation}
where $\pi_i$ is the probability that $\tilde{y}_i = 1$ given the true model parameters, and $\pi_{\lambda,i}$ is the corresponding quantity under the $\lambda$-sparsified model. This expression for the logistic excess error relates to the linear model case in that each expression can be derived from
\begin{equation}
\psi_{\lambda} = \sqrt{n^{-1} \E\left(||\tilde{Y} - \hat{Y}_{\lambda}||^2\right)} - \sqrt{n^{-1} \E\left(||\tilde{Y} - \E(\tilde{Y})||^2\right)}
\end{equation}
where the expectation is with respect to the predictive distribution of $\tilde{Y}$ conditional on the model parameters, and $\hat{Y}_{\lambda}$ denotes the optimal $\lambda$-sparse prediction. In particular, $\hat{Y}_{\lambda} \equiv \X\beta_{\lambda}$ for the linear model and $\hat{y}_{\lambda,i} \equiv \pi_{\lambda,i} = (1 + \exp{- X_i\beta_{\lambda})^{-1}}$ for the logistic regression model. One notable difference between the expressions for excess error under the linear model and the logistic model is that the linear model has constant variance whereas the variance term depends on the predictor point in the logistic model as a result of the Bernoulli likelihood.

\subsection*{Example: German credit data ($n =1000$, $p = 48$)}
\begin{figure}[h!]
\begin{center}
\includegraphics[width=6.5in]{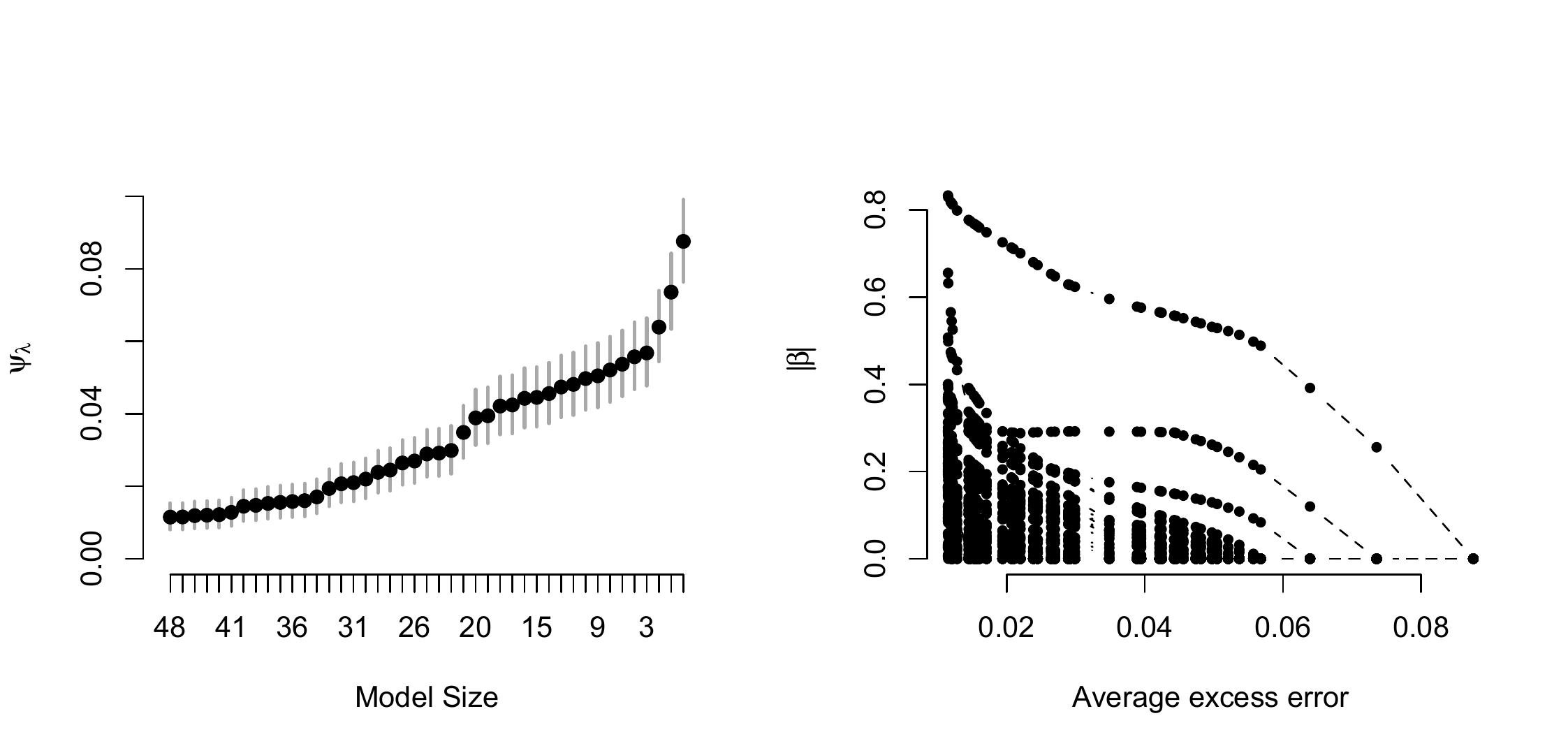}
\caption{DSS plots for the German credit data.  For this data, each included variable seems to add an incremental amount, as the excess error plot builds steadily until reaching the null model with no predictors.}\label{german}
\end{center}
\end{figure}

To illustrate selection summary in the logistic regression context, we use the German Credit data from the UCI repository, where $n=1000$ and $p=48$. In each record we have available covariates associated with a loan applicant, such as credit history, checking account status, car ownership and employment status.   The outcome variable is a judgment of whether or not the applicant has ``good credit".  A natural objective when analyzing this data would be to develop a good model for assessing creditworthiness of future applicants.  A default shrinkage prior over the regression coefficients is used, based on the ideas described in \cite{BayesLogit} and the associated {\tt R} package {\tt BayesLogit}. The DSS selection summary plots (adapted to a logistic regression) are displayed in Figure \ref{german}. The plot suggests a high degree of ``pre-variable selection", in that all of the predictor variables appear to add an incremental amount of prediction accuracy, with no single predictor appearing to dominate.  Nonetheless, several of the larger models (smaller than the full forty-eight variable model) do not give up much in excess error, suggesting that a moderately reduced model ($\approx 35$), may suffice in practice.  Depending on the true costs associated with measuring those ten least valuable covariates, relative to the cost associated with an increase of 0.01 in excess error, this reduced model may be preferable.

\subsection{Selection summary for Gaussian graphical models}
Covariance estimation is yet another area where a sparsifying loss function can be used to induce a parsimonious posterior summary.  

Consider  a $(p\times1)$ vector $(x_1, x_2,\dots,x_p) = \mathbf{X} \sim \mbox{N}(0,\mathbf{\Sigma})$. Zeros in the precision matrix $\mathbf{\Omega} = \mathbf{\Sigma}^{-1}$ imply conditional independence among certain dimensions of $\mathbf{X}$. As sparse precision matrices can be represented through a labelled graph, this modeling approach is often referred to as Gaussian graphical modeling.  Specifically, for a graph $G=(V,E)$, where $V$ is the set of vertices and $E$ is the set of edges, let each edge represent a non-zero element of $\mathbf{\Omega}$.  See \cite{Jones05} for a thorough overview. This problem is equivalent to finding a sparse representation in $p$ separate linear models for $X_j | X_{-j}$, making the selection summary approach developed above directly applicable.

As with linear models, one has the option of modeling the entries in the precision matrix via shrinkage priors or via selection priors with point masses at zero. Regardless of the specific choice of prior, summarizing patterns of conditional independence favored in the posterior distribution remains a major challenge. 

A DSS parsimonious summary can be achieved via a multivariate extension of (\ref{Loss}) by once again leveraging the notion of  ``predictive accuracy'' as defined by the negative log likelihood:
\begin{equation}
\mathcal{L}(\tilde{\mathbf{X}}, \mathbf{\Gamma}) = \lambda ||\mathbf{\Gamma}||_0 - \log \det(\mathbf{\Gamma}) - \mbox{tr}(n^{-1}\tilde{\mathbf{X}} \tilde{\mathbf{X}^\prime} \mathbf{\Gamma}) 
\end{equation}
where  $\mathbf{\Gamma}$ represents the decision variable for $\mathbf{\Omega}$ and $||\mathbf{\Gamma}||_0$ represents the sum of non-zero entries in off-diagonal elements of $\mathbf{\Gamma}$. Taking expectations with respect to the posterior predictive of $\mathbf{\tilde{X}}$  yields
\begin{equation}
\mathcal{L}(\mathbf{\Gamma}) =  \E\left( \mathcal{L}(\tilde{\mathbf{X}}, \mathbf{\Gamma})\right)  = \lambda ||\mathbf{\Gamma}||_0 - \log \det(\mathbf{\Gamma}) - \mbox{tr}(\mathbf{\bar{\Sigma}} \mathbf{\Gamma}) 
\end{equation}
where $\mathbf{\bar\Sigma}$ represents the posterior mean of $\mathbf{\Sigma}$.

As before, an approximate solution to the DSS graphical model posterior summary optimization problem can be obtained by employing the surrogate $\ell_1$ penalty
\begin{equation}
\mathcal{L}(\mathbf{\Gamma}) =  \E\left( \mathcal{L}(\tilde{\mathbf{X}}, \mathbf{\Gamma})\right)  = \lambda ||\mathbf{\Gamma}||_1 - \log\det(\mathbf{\Gamma}) - \mbox{tr}(\mathbf{\bar{\Sigma}} \mathbf{\Gamma}). 
\end{equation}
as developed by penalized likelihood methods such as the graphical lasso \citep{GLASSO}.

\end{document}